\documentclass[]{BioPhysMath}
\usepackage[OT4,T1]{fontenc}
\usepackage[utf8]{inputenc}
\usepackage{polski}
\usepackage{graphicx}
\usepackage{float}
\usepackage{amsmath}
\usepackage{amssymb}
\usepackage{filecontents}

\usepackage[english,polish]{babel} 
\usepackage{lastpage} 

\usepackage{etex} 

\usepackage{color} 
\usepackage{cite} 

\RequirePackage[numbers]{natbib}

\usepackage[colorlinks=true]{hyperref}
  \hypersetup{
    pdftitle={Template BioPhysMath}, 
    pdfauthor=Nowy Autor, 
    colorlinks,
    urlcolor=blue,
    filecolor=magenta,
    citecolor=green, 
    linkbordercolor={1 1 1}, 
    citebordercolor={1 1 1},  
    urlbordercolor={ 1 1 1}  
  } 

\newcommand{\orcid}[1]{\href{https://orcid.org/#1}{\includegraphics[scale=.05]{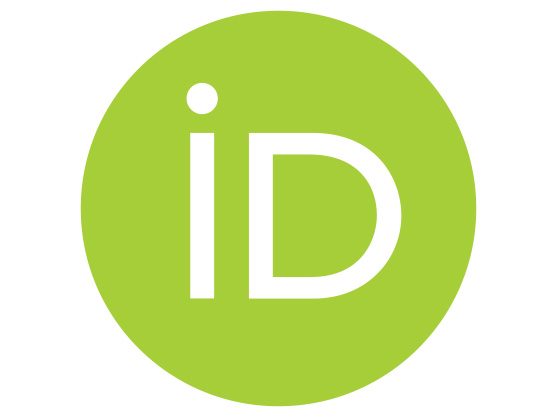}}}

\pdfoutput=1
\usepackage{graphicx}
\usepackage{wrapfig}
\graphicspath{{./Figures/},{./Pictures/}}
\usepackage{multicol}
\firstpage{i}    
\LogoG{\vspace*{1cm}\includegraphics[width=0.1\textwidth]{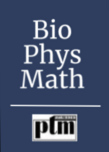}}
\volume{1}
\fasc{1}
\years{2024}
\LogoGMS{\includegraphics[width=0.18\textwidth]{bpm.png}}
\volumeMS{1}
\numberMS{1}
\wwwtrue

\pages{9}
\received[9th February 2024]{7th April 2023}
\lastrevision{9th February 2024}
\logo{}{}{}
\def\doinum{10.14708/bpm.v43i1.xxx}

\title[Phase transitions in the Prisoner's Dilemma game]{Phase transitions in the Prisoner's Dilemma game on scale-free networks}

\tpauthortrue 
\author[Jacek Mi\c{e}kisz]{Jacek Mi\c{e}kisz\orcid{0000-0001-6651-8951}}
\thanks{Corresponding author}
\affiliation{Institute of Applied Mathematics and Mechanics, University of Warsaw}
\address{Banacha 2, 02-097 Warsaw, Poland}
\email{miekisz@mimuw.edu.pl}

\author[Javad Mohamadichamgavi]{Javad Mohamadichamgavi\orcid{0000-0001-7208-5996}}
\affiliation{Institute of Applied Mathematics and Mechanics, University of Warsaw}
\address{Banacha 2, 02-097 Warsaw, Poland}
\email{jmohamadi@mimuw.edu.pl}

\author[Jakub \L\c{a}cki]{Jakub \L\c{a}cki\orcid{}}
\affiliation{Institute of Applied Mathematics and Mechanics, University of Warsaw}
\address{Banacha 2, 02-097 Warsaw, Poland, now at Google Research, New York, USA}
\email{j.lacki@mimuw.edu.pl}


\begin{document}
\vspace{-5ex}
\renewcommand{\thefootnote}{}
\footnote{\href{http://creativecommons.org/licenses/by/3.0/}{Licensed under a Creative Commons Attribution License (CC-BY)}}
\setcounter{page}{1} 
\selectlanguage{english}\Polskifalse

\begin{abstract}
We study stochastic dynamics of the Prisoner's Dilemma game on random Erd\"{o}s-R\'{e}nyi and Barab\'{a}si-Albert networks 
with a cost of maintaining a link between interacting players. Stochastic simulations show that when the cost increases, 
the population of players located on Barab\'{a}si-Albert network undergoes a sharp transition from an ordered state, where almost all players cooperate, 
to a state in which both cooperators and defectors coexist. At the critical cost, the population oscillates in time between these two states.
Such a situation is not present in the Erd\"{o}s-R\'{e}nyi network. We provide some heuristic analytical arguments for the phase transition 
and the value of the critical cost in the Barab\'{a}si-Albert network.
\end{abstract}

\section{Introduction}

Cooperation between unrelated individuals in human and animal societies is an intriguing issue 
in biology and social sciences \cite{hamilton,hammer,axelrod,nowakbook1,nowakbook2,sigmundbook}. 
One can describe it within the framework of evolutionary game theory and especially the Prisoner's Dilemma game. 
In this game, two players simultaneously decide whether to cooperate or to defect. The mutual cooperation gives both of them the reward $R$ 
which is higher than the punishment $P$ resulting from the mutual defection. However, a cooperating player is tempted to defect 
to receive the highest payoff $T$ leaving the other cooperating player with the lowest payoff $S$. 
Payoff inequalities $S<P<R<T$ imply that defection gives a player a higher payoff than cooperation regardless of a strategy adopted by its opponent. 
Therefore rational individuals defect in spite of the fact that they would be better off if they cooperated.  

In the framework of evolutionary game theory \cite{maynard,weibull,hof2}, payoffs are interpreted as numbers of offspring who inherit strategies 
of their parents. The evolution of very large (infinite) populations is usually modeled by differential or difference replicator equations
which describe time changes of fractions of the population of individuals playing given strategies \cite{tayjon,hof1}. 
In the case of the Prisoner's Dilemma, the long-run of such dynamics is the population consisting of just defectors.

In replicator dynamics, players receive average payoffs weighted by frequencies of strategies in the infinite population. 
However, real populations are finite and individuals receive payoffs (not average payoffs) which result from interactions with random opponents 
in well-mixed populations or neighbors in spatially structured populations. In their pioneering paper \cite{nowak0}, Nowak and May located players 
on regular graphs and allowed them to interact only with their neighbors. The payoff of any player is then the sum of payoffs 
resulting from individual games. In discrete time moments, players imitate neighbors with the highest payoff obtained in the previous round, 
making perhaps mistakes. In stationary states of such stochastic dynamics, various structures of coexisting cooperators 
and defectors were observed \cite{nowak1,nowak2}. Since then various versions of spatial Prisoner's Dilemma and other games have been extensively studied, 
see a review paper \cite{szaboreview}. It was shown and generally understood that cooperation can be maintained 
in space-structured populations. Cooperating players tend to form clusters, receive high payoffs and therefore are immune to invasion by defectors. 
Recently there appeared papers indicating that the structure of a network on which players are located may play a significant role in promoting the cooperation. 
Various non-regular and random graphs were investigated. In particular, Santos and Pacheco \cite{santos1,santos2} 
shown that the scale-free Barab\'{a}si-Albert network favors cooperation for a large range of game parameters. 

In such a heterogeneous graph, there are vertices with many edges, the so-called hubs. Players located on hubs interact with many individuals. 
It was shown that existence of hubs favors cooperation. However, maintaining social ties can be costly. It is natural therefore to introduce 
participation costs in spatial games. It was shown in \cite{masuda} that participation costs reduce the advantage of heterogeneous networks 
in maintaining a high level of cooperation. 

Here we study the equilibrium behavior of the imitation dynamics of systems of interacting individuals playing the Prisoner's Dilemma game 
on random Erd\"{o}s-R\'{e}nyi \cite{er} and Barab\'{a}si-Albert networks \cite{ba1,ba2}. The stochastic dynamics in spatial games are
similar to stochastic updating in the Ising and lattice-gas models in statistical mechanics. However, in spatial games, in general there does not exist 
a global order parameter, like the energy or free energy in the Ising model, which the system wants to optimize. Similarities and differences 
between stochastic dynamics in spatial games and in systems of many interacting particles were discussed in \cite{blume,cime,szaboreview}. 

Critical phenomena in random networks were studied very extensively, for a review see \cite{dorogov}, mean-field approximation in the Ising model 
on the Barab\'{a}si-Albert network was used in \cite{holyst,ginestra}, phase transitions in voter models were analysed in \cite{sznajd,voter1,voter2}.
We have performed Monte-Carlo simulations to explore dependence of the cooperation level in the stationary state of the imitation dynamics 
on the participating cost. We report that in the case of the Barab\'{a}si-Albert network we observe a critical value of the cost 
at which a population changes abruptly from a high to a lower level of cooperation. This follows up constructions and simulations presented in \cite{sulkowski}

\section{Model}
\label{sec:2}

Players are located on vertices of the Erd\"{o}s-R\'{e}nyi (ER) \cite{er} and the Barab\'{a}si-Albert (BA) networks \cite{ba1,ba2}.
We build the ER network by putting with probability $p$ an edge between every pair of $N=10^{4}$ vertices. 
It follows that the average degree of vertices (the average number of neighbors) is equal to $\alpha=p(N-1).$ 
The BA network is built by the preferential attachment procedure. 
We start with $m_{o}$ fully connected vertices and then we add $N-m_{o}$ vertices, each time connecting them with $m$ already available vertices 
with probabilities proportional to their degrees. If $m_{o}=\alpha+1$ and $m=\alpha/2$, then we get a graph with the average degree equal to $\alpha$. 
It is known that such a graph is scale-free with the probability distribution of degrees given by $p(k)\sim k^{-3}$ \cite{ba1,ba2,durett}.

Individuals play with their neighbors the Prisoner's Dilemma game. For a general Prisoner's Dilemma game 
we introduce a costs $\gamma$ of maintaining a link payed by both connected players \cite{masuda} and hence our payoff matrix reads:

\[
\begin{array}{ccc}
  & C & D \\
\noalign{\medskip}
C & 1-\gamma & S-\gamma \\
\noalign{\medskip}
D & T-\gamma & P-\gamma
\end{array}
\]

where the entry $ij$ is the payoff of the row player using the $i$-th strategy while the column player uses the $j$-th one.  

At discrete moments of time, all individuals interact with their neighbors and receive payoffs which are sums with respect to individual games. 
Then the imitation process takes place. A randomly chosen player compares its payoff to payoffs of all its neighbors 
and with the probability $1-\epsilon$ chooses the strategy which provided the highest payoff in the previous round 
and with the probability $\epsilon$ adopts the other one, we fix $\epsilon=10^{-3}$. We interpret $\epsilon$ as a measure 
of irrationality of players or simply the noise level.
This completes one step of the discrete-time dynamics - a Markov chain with $2^{N}$ states. The presence of noise makes our chain irreducible  - 
any state can be reached from any other state in a finite number of steps. It is also aperiodic which follows from the fact that the chain can remain at any state 
with a non-zero probability. This means that our Markov Chain is ergodic - it has a unique stationary probability distribution, 
the so-called stationary state, to which any initial probability distribution converges.

To find a cooperation level in the stationary state we perform stochastic simulations.
We start with a completely random initial conditions with the fraction 
of cooperators $=1/2$. Then we perform $10^{5}$ Monte-Carlo rounds followed by $10^{4}$ rounds, 
in which frequencies of cooperators are computed. One round consists of $N=10^4$ steps, where $N$ is the number of players, 
so that in every round, on average each player has the opportunity to update its strategy. We repeat simulations 50 times and average the results.

\section{Results}

\begin{figure}[H]
\centering
\includegraphics[scale=0.35]{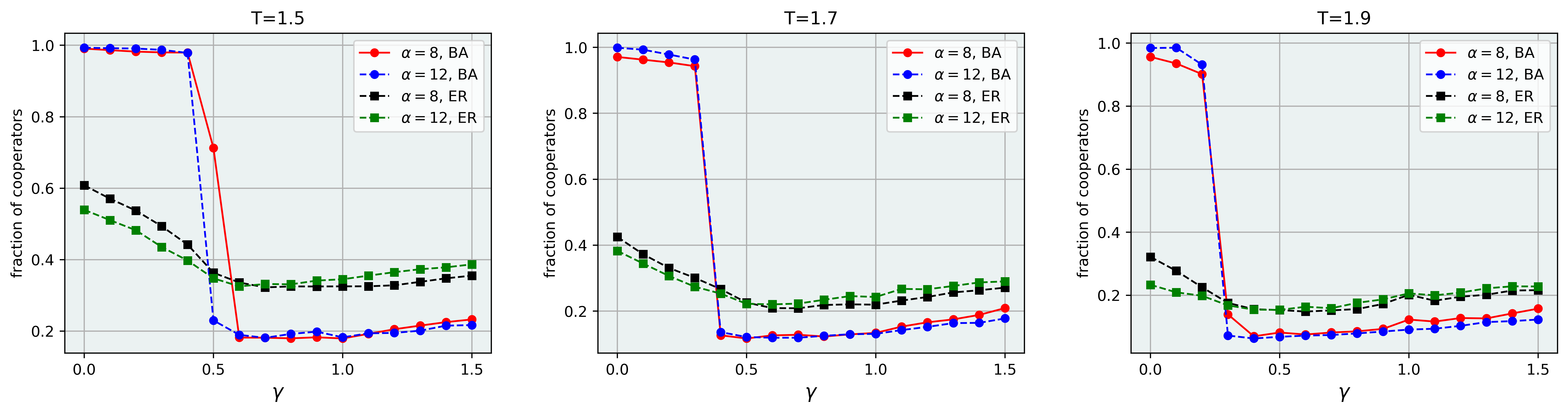}
 \caption{\label{graph_gamma}Fraction of cooperators in the stationary state as a function of a cost of maintaining a link.}
\end{figure}

\begin{figure}[H]
\centering
\includegraphics[scale=0.40]{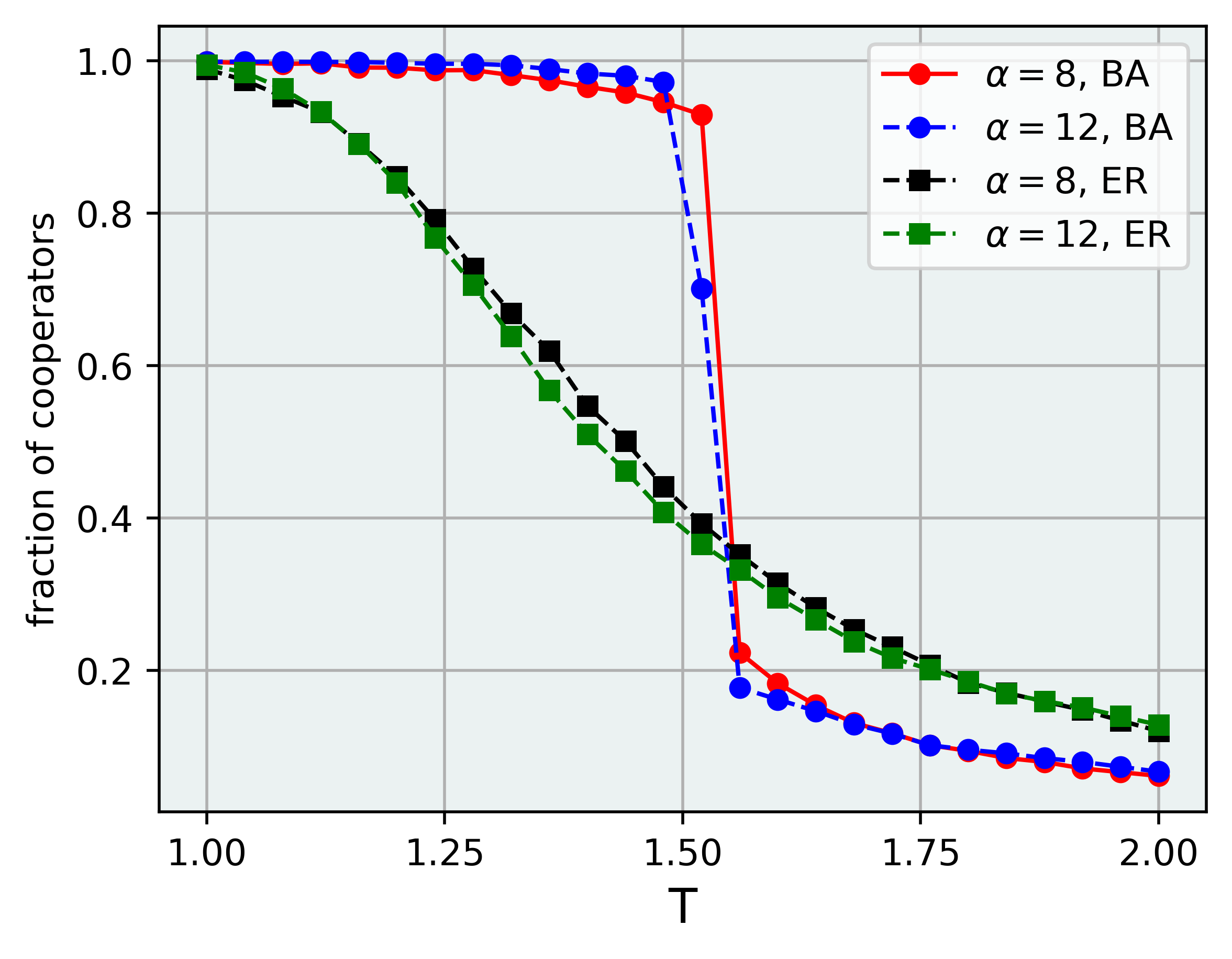}
 \caption{\label{graph_t} Fraction of cooperators in the stationary state as a function of $T$, $\gamma = 0.46$.}
\end{figure}

\begin{figure}[H]
\centering
\includegraphics[scale=0.35]{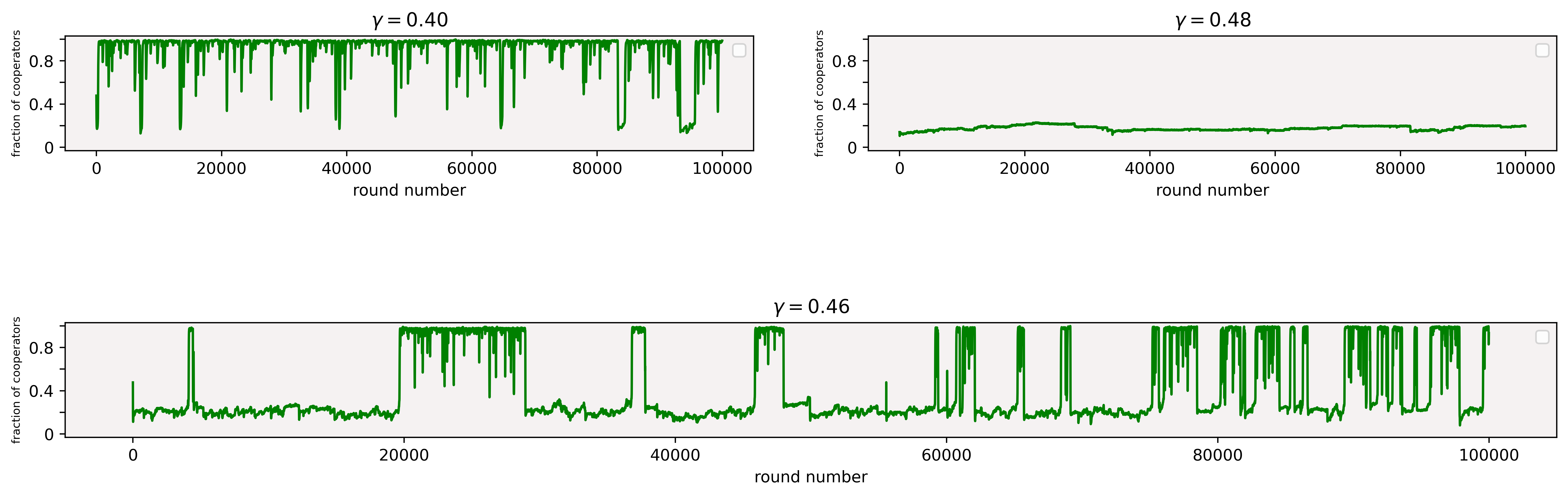}
 \caption{\label{after_every_turn} Fraction of cooperators after each round in a sample simulation for various values of $\gamma$. 
Barab\'{a}si-Albert network, $T=1.5$, $\alpha = 12$.}
\end{figure}

It is usually assumed, see for example \cite{nowak0,santos1}, that $S=P=0, R=1$  
and $T>1$, such a game is called a weak Prisoner's Dilemma. We will follow this practice here.
Stationary fractions of cooperators for various average degrees of vertices $\alpha$ of ER  
and BA networks as a function of the cost $\gamma$ of maintaining one link, for the temptation to defect $T=1.5, 1.7$, and $1.9$ 
are shown in Fig. \ref{graph_gamma} and as a function of $T$ for $\gamma=0.46$ in Fig. \ref{graph_t}. We observe that the cost $\gamma$ plays the crucial role 
in the long-run behavior of our systems. The effect of $\gamma$ is much more pronounced for the BA network than for the ER one. 
For small positive values of $\gamma$, the level of cooperation is much higher for the BA network than for the ER one; 
for bigger $\gamma$ the cooperation level is higher for the ER network. 
Our main result is that in the case of the BA network, when the cost increases, the population of players undergoes a sharp transition 
from an efficient ordered state, where almost all players cooperate, to a disordered state in which both cooperators and defectors coexist.
For $T=1.5$, this critical value of $\gamma$ is about $0.46$ as is seen in Fig. 1. This is reminiscent of the first-order discontinuous phase transition present 
in statistical mechanics models of interacting particles. In such models, at the critical point, there coexist two (or more) phases of the system. 
A typical example is the presence of two phases, up and down, in the ferromagnetic Ising model at the zero external magnetic field 
below the critical Curie temperature. To see if such a situation may be present here in the model of interacting players, 
we looked at the time evolution of the frequency of cooperation. In Fig. \ref{after_every_turn} we see that for $\gamma = 0.4$
($T$=1.5 and $\alpha=12$), that is below a critical value, the population basically stays at an ordered state where almost all players cooperate, 
for $\gamma = 0.48$, the population settles at a state in which both cooperators and defectors coexist. 
For $\gamma = 0.46$, we observe that the system oscillates between these two states.
Again, this is a typical situation in finite systems of interacting particles with a discontinuous phase transition in the infinite-system limit.
Spatial structure favors cooperation because clusters of cooperators have relatively high payoffs and therefore are resistant to attacks of defectors 
\cite{nowak0}.
So it is natural to expect that the cooperation level is a decreasing function of the linking cost \cite{masuda}.
To understand the existence of an abrupt decrease of the cooperation level in the Barab\'{a}si-Albert network, we calculate the linking cost
at which an average payoff of $D-$strategy is bigger than the average payoff of a $C-$strategy in the neighborhood of a solitary $D-$player.
More precisely, let us assume that we have a single $D-$player in the population of cooperators. 
Its average payoff is equal to the expected value of the vertex degree, $<k>=\alpha$ multiplied by $T-\gamma$.
Now choose one of its $C-$neighbors. It is known that its degree distribution is proportional to $kp(k)$, where $p(k)$ 
is the degree distribution of the network \cite{feld,newman}. Therefore the expected value of its degree is given by $<k^2>/<k>$.
We set the expected value of the $D-$player to be equal to the expected value of its neighbor - the $C-$ player, that is

\begin{equation}
<k>(T-\gamma)=(1-\gamma)(\frac{<k^2>}{<k>}-1)-\gamma.
\end{equation}

We solve the above equation for $\gamma$ and get

\begin{equation}
\gamma=\frac{\frac{<k^2>}{<k>}-<k>T-1}{\frac{<k^2>}{<k>}-<k>}.
\end{equation}

For the expected value of the square of vertex degrees we use the following formula \cite{K2}:

\begin{equation}
<k^2> =<k>^2 \frac{\ln N}{4}.
\end{equation}

For $<k>=\alpha=8$ and $N=10000$ we obtain critical values of the cost equal aproximately to $0.519$ for $T=1.5$ and $0.21$
for $T=1.9$ which is in a good agreement with values obtained in simulations as can be seen in Fig. 1. Further more rigorous analysis is needed.

\begin{figure}[H]
\centering
\includegraphics[scale=0.4]{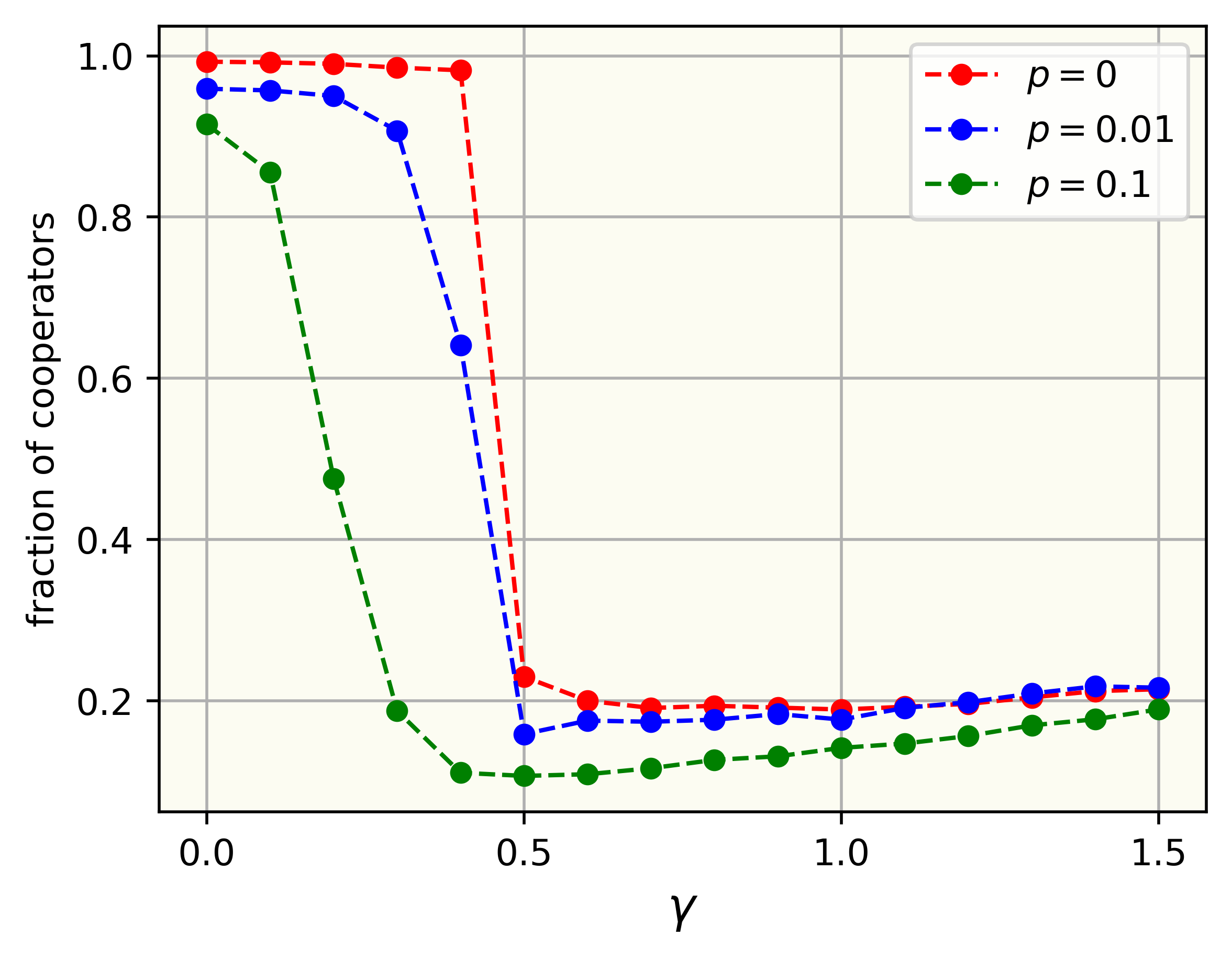}
 \caption{\label{fig4}Fractions of cooperators depending on the cost of maintaining a link for different values of $P$. 
Barab\'{a}si-Albert network, $T=1.5$, $\alpha = 12$.}
\end{figure}

Let us now consider a standard Prisoner's Dilemma game to examine how changing the payoff for mutual defection, $P$, while holding other payoff values constant, 
affects the level of cooperation. We fix payoffs as $T=1.5, R=1$, and $S=0$, 
we vary $\gamma$ and observe its effect on cooperation for various $P's$ in the Barab\'{a}si-Albert network. 
The results in Fig. \ref{fig4} demonstrate that as $P$ increases, the critical $\gamma$ value, where a phase transition occurs, decreases. 
This is because, with a larger $P$, defectors obtain higher payoffs when playing each other, reducing the incentive to cooperate for any fixed $\gamma$.

\section{Discussion} 

We investigated how the cost of maintaining links between players affects the cooperation level in the spatial Prisoner's Dilemma games. 
In the case of the Barab\'{a}si-Albert network, we observed that when the cost increases, the population of players undergoes a sharp transition 
from a high to a lower level of cooperation. Our numerical simulations of the time evolution of the frequency of cooperation showed that at the critical cost 
the population oscillates between two states. It means that at such a cost there coexist two population states: an ordered one 
where almost all players cooperate and one in which both cooperators and defectors coexist. 
We provided some heuristic analytical arguments for the existence of a phase transition and the value of the citical cost. 

Further research is needed to elucidate the nature of this phase transition and to provide exact or at least an approximate analytical analysis of the observed behavior.

\vskip 0.5cm
\acknowledgments{ This project has received funding from the European Union’s Horizon 2020 research and innovation programme 
under the Marie Sk\l odowska-Curie grant agreement No 955708. J. Mi\c{e}kisz would like to thank Bartosz Su\l kowski, constructions and simulations contained in his bachelor thesis written in 2005 have been extended and are presented in this paper. Computer simulations were made with the support 
of the Interdisciplinary Center for Mathematical and Computational Modeling of the University of Warsaw (ICM UW).}
\vskip 0.5cm

\centerline{\large\bf References}


\begin{thebibliography}{99}
\bibitem{hamilton} W. D. Hamilton, The evolution of altruistic behavior, Am. Nat. \textbf{97}, 354-356 (1963).
\bibitem{hammer} P. Hammerstein, ed. \emph{Genetic and cultural evolution of cooperation}, MIT press (2003).
\bibitem{axelrod} R. M. Axelrod, \emph{The Evolution of Cooperation}, Basic Books, New York (1984)
\bibitem{nowakbook1} M. A. Nowak, \emph{Evolutionary Dynamics: Exploring the Equations of Life}, Belknap Press of Harvard University Press (2006).
\bibitem{nowakbook2} M. Nowak and R. Highfield, \emph{SuperCooperators: Altruism, Evolution, and Why We Need Each Other to Succeed}, Free Press (2011).
\bibitem{sigmundbook} K. Sigmund, \emph{The Calculus of Selfishness}, Princeton University Press (2010).
\bibitem{maynard} J. Maynard Smith, \emph{Evolution and the Theory of Games}, Cambridge University Press (1982).
\bibitem{weibull} J. W. Weibull, \emph{Evolutionary Game Theory}, The MIT Press (1995).
\bibitem{hof2} J. Hofbauer and K. Sigmund, \emph{Evolutionary Games and Population Dynamics}, Cambridge University Press (1998)
\bibitem{tayjon} P. Taylor and L. B. Jonker, Evolutionary stable strategies and game dynamics, Math. Biosci. \textbf{40}, 145-156 (1978).
\bibitem{hof1} J. Hofbauer, P. Schuster, and K. Sigmund, A note on evolutionary stable strategies and game dynamics, 
J. Theor. Biol. \textbf{81}, 609-612 (1979).
\bibitem{nowak0} M. A. Nowak and R. M. May, Evolutionary games and spatial chaos, Nature \textbf{359}, 826-829 (1992).
\bibitem{nowak1} M. A. Nowak and R. M. May, The spatial dilemmas of evolution, Int. J. Bifurcat. Chaos \textbf{3}, 35-78 (1993).
\bibitem{nowak2} M. A. Nowak, S. Bonhoeffer, and R. M. May, More spatial games, Int. J. Bifurcat. Chaos \textbf{4}, 33-56 (1994).
\bibitem{szaboreview} G. Szab\'{o} and G. F\'{a}th, Evolutionary games on graphs, Phys. Rep. \textbf{446}, 97-216 (2007).
\bibitem{santos1} F. C. Santos and J. M. Pacheco, Scale-free networks provide a unifying framework for the emergence of cooperation,  Phys. Rev. Lett. \textbf{95}, 098-104 (2005).
\bibitem{santos2} F. C. Santos, J. Rodrigues, and J. M. Pacheco, Graph topology plays a determinant role in the evolution of cooperation, P. Roy. Soc. B. \textbf{273}, 51-55 (2006).
\bibitem{masuda} N. Masuda, Participation costs dismiss the advantage of heterogeneous networks in evolution of cooperation, Proc. Roy. Soc. B. \textbf{274}, 1815-1821 (2007).
\bibitem{er} P. Erd\"{o}s and A. R\'{e}nyi, On random graphs I, Publ. Math. Debrecen, \textbf{6}, 290-297 (1959).
\bibitem{ba1} A. L. Barabasi and R. Albert, Emergence of scaling in random networks, science \textbf{286}, 509-512 (1999).
\bibitem{ba2} R. Albert and A. L. Barabasi, Statistical mechanics of complex networks, Rev. Mod. Phys. \textbf{74}, 47-97 (2002).
\bibitem{blume} L. E. Blume, The statistical mechanics of strategic interaction, Games Econ. Behav. \textbf{5}, 387-424 (1993).
\bibitem{cime} J. Mi\c{e}kisz, Evolutionary game theory and population dynamics, in Multiscale Problems in the Life Sciences, From Microscopic to Macroscopic, 
V. Capasso and M. Lachowicz (eds.), Lecture Notes in Mathematics \textbf{1940}, 269-316 (2008).
\bibitem{dorogov} S. N. Dorogovtsev, A. V. Goltsev, and J. F. F. Mendes, Critical phenomena in complex networks, Rev. Mod. Phys. \textbf{80}, 1275-1335 (2008).
\bibitem{durett} R. Durrett, \emph{Random Graph Dynamics}, Cambridge University Press (2006).
\bibitem{holyst} A. Aleksiejuk, J. A. Ho\l yst, and D. Stauffer, Ferromagnetic phase transition in Barab\'{a}si-Albert networks, Physica A \textbf{310}, 260-266 (2002).
\bibitem{ginestra} G. Bianconi, Mean field solution of the Ising model on a  Barab\'{a}si–Albert network, Physics Letters A \textbf{303}, 166-168 (2002).
\bibitem{sznajd} A. Chmiel and K. Sznajd-Weron, Phase transitions in the q-voter model with noise on a duplex clique, Phys. Rev. E \textbf{92}, 052812 (2015).
\bibitem{voter1}A. J\c{e}drzejewski, Pair approximation for the q-voter model with independence on complex networks, Phys. Rev. E \textbf{95}, 012307 (2017).
\bibitem{voter2} A. J\c{e}drzejewski, J. Toruniewska, K. Suchecki, O. Zaikin, and J. A. Ho\l yst, 
Spontaneous symmetry breaking of active phase in coevolving nonlinear voter model, Phys. Rev. E \textbf{102}, 042313 (2020).
\bibitem{sulkowski} B. Su\l kowski,  bachelor thesis, University of Warsaw, in Polish  (2005).
\bibitem{feld} S. Feld, Why your friends have more friends than you do, American Journal of Sociology \textbf{96}, 1464–1477 (1991).
\bibitem{newman}M .E. J. Newman, S. H. Strogatz, and D. J.  Watts, Random graphs with arbitrary degree distributions and their
 applications. Phys. Rev. E \textbf{64}, 026118 (2021).
\bibitem{K2} A.P. Alodjants, A. Yu. Bazhenov, and M. M. Nikitina, Phase transitions in quantum complex networks, J. Phys.: Conf. Ser. \textbf{2249} 012014 (2022).
\end{thebibliography}
\bibliographystyle{abbrv}
\Koniec
\end{document}